\documentstyle[preprint,prl,aps]{revtex}
\preprint{IMSc-97/12/38}

\def\Im{{\rm Im}}
\def\Re{{\rm Re}}
\def\dsp{\displaystyle}
\def\nn{\nonumber}

\begin{document}
\title{Determination of the angle $\gamma$ using 
$B\to D^*V$ modes}
\author{Nita Sinha and Rahul Sinha
\ifpreprintsty
\footnote{e-mail: nita@imsc.ernet.in,sinha@imsc.ernet.in}
\else
\cite{author}
\fi}
\address{ Institute of Mathematical Sciences, Taramani, Chennai 600113, 
India.}
\date{\today}
\maketitle
\begin{abstract}            

We propose a method to determine the angle $\gamma=arg(V_{ub})$, using
the $B\to D^*V$ ($V=K^*, \rho$) modes. The $D^*$ is considered to
decay to $D \pi$. An interference of the $B \to D^{*0}V$ and $B \to
\overline {D^{*0}}V$ amplitudes is achieved by looking at a common
final state $f$, in the subsequent decays of $D^0/\overline{D^0}$. A
detailed analysis of the angular distribution, allows determination,
not only of $\gamma$ and $|V_{ub}|$, but also all the hadronic
amplitudes and strong phases involved. No prior knowledge of doubly
Cabibbo suppressed branching ratios of $D$ are required. Large $CP$
violating asymmetries ($\sim30 \%$ for $\gamma=30^o$) are possible if
$\overline{D^0} \to f$ is doubly Cabbibo suppressed, while $D^0 \to f$
is Cabbibo allowed, for decays of $B^+$ or $B^0$.
\end{abstract}

$CP$ violation is one of the unsolved mysteries in particle
physics. In the standard model, however, it is parameterized by
including a phase in the unitary Cabbibo--Kobayashi--Maskawa (CKM) 
matrix\cite{KM}. The aim of the several upcoming factories and 
detectors dedicated to studying $B$ physics is to test this 
parameterization, by measuring the three angles 
of the unitarity triangle \cite{Quinn}. The angle $\gamma$, which is the 
phase of the element $V_{ub}$ of the CKM matrix, is one of the most 
difficult to measure\cite{Quinn}. $\gamma$ is also important, as its 
non-vanishing value is a signal of direct $CP$ violation. Though $CP$ 
violation was seen in $K$ system, more than 30 years ago, no signature of 
direct $CP$ violation has yet been established. 

One of the promising methods of measuring the angle $\gamma$ is the so
called GLW method\cite{GLW,Dunietz}. In this method $\gamma$ is
obtained from an interference of the mode $B\to D^0K$ with
$B\to\overline{D^0}K$, which occurs if and only if, both $D^0$ and
$\overline{D^0}$ decay to a common final state $f$; in particular, $f$
is taken to be a $CP$ eigenstate. This technique of extracting
$\gamma$ requires a measurement of the branching ratio for $B^+\to D^0
K^+$ which is not experimentally feasible as pointed out in
\cite{ADS}. Moreover, the $CP$ violating asymmetries tend to be small
as the interfering amplitudes are not comparable. The use of non--$CP$
eigenstates `$f$' has also been considered \cite{Dunietz-GLW} in
literature. Recently Atwood, Dunietz and Soni (ADS)\cite{ADS} extended
this proposal by considering `$f$' to be non--$CP$ eigenstates that
are also doubly Cabbibo suppressed modes of $D$. The two interfering
amplitudes then are of the same magnitude resulting in large
asymmetries. Their proposal is to use two final states $f_1$ and $f_2$
with at least one being a non--$CP$ eigenstate. The use of more than
one final state enables not only the determination of $\gamma$, but
also of all the strong phases involved and the difficult to measure
branching ratio $Br(B^+\to D^0 K^+)$. However, an input into the
determination of $\gamma$ is the branching ratio of the doubly Cabbibo
suppressed mode of $D$.  Though $D$ decays have been studied for a
long time, only one doubly Cabibbo suppressed mode has been observed
with an error that is currently as large as $50\%$.

In this letter we extend these proposals to the corresponding decays of 
$B$ into two vector mesons, by considering $B\to D^*V$, where $V$ is 
either a $K^*$ or $\rho$. The $ D^{*0} /\overline{D^{*0}}$ will decay 
into $D^0/\overline{D^0}$, which if subsequently decays to a final state 
`$f$' that is common to both $D^0$ and $\overline{D^0}$, then the two 
decay channels $D^{*0}V$ and $\overline{D^{*0}}V$ can interfere, giving 
rise to the desired $CP$ violating effects. The several amplitudes
provided by the various partial waves of a single vector--vector final 
state, enable us to extract $\gamma$, all the relevant hadronic 
amplitudes and strong phases, thereby removing  any hadronic 
uncertainties. Our approach does not require a prior knowledge of the 
poorly known doubly Cabibbo suppressed branching ratios of $D$, which
in fact can be determined here, due to interference effects.  

The most general covariant amplitude for a $B$ meson decaying to a pair of 
vector mesons has the form \cite{Valencia,Krammer}: 
\begin{equation} 
A(\displaystyle B(p)\to V_1(k) V_2(q))= 
\displaystyle\epsilon_1^{*\mu}\epsilon_2^{*\nu} \left( a g_{\mu\nu}+\nn\\ 
\frac{b}{m_1 m_2} p_{\mu} p_{\nu}+i\frac{c}{m_1 m_2} 
\epsilon_{\mu\nu\alpha\beta} k^\alpha q^\beta \right), 
\end{equation} 
where, $\epsilon_1$, $\epsilon_2$ and $m_1$,$m_2$ 
represent the polarization vectors and the masses of the vector mesons 
$V_1$ and $V_2$ respectively. The coefficients $a,b$ and $c$ can be 
expressed in terms of the linear polarization basis $A^\|$, $A^0$ and 
$A^\perp$\cite{Krammer}. If both the vector mesons subsequently decay to two 
pseudoscalar mesons, {\it i.e.} $V_1\to P_1 P_1^\prime$ and $V_2 \to P_2 
P_2^\prime$, the amplitude can be expressed as,
\begin{equation} 
A(\displaystyle B\to V_1 V_2)=4|\vec{k_1}||\vec{q_1}|\left(-A^0 
\cos\theta_1 \cos\theta_2 -\frac{A^\|}{\sqrt{2}}\sin\theta_1 
\sin\theta_2 \cos\phi +i \frac{A^\perp}{\sqrt{2} }\sin\theta_1 
\sin\theta_2 \sin\phi\right), 
\end{equation} 
where, $\theta_1$ ($\theta_2$) is the angle between the $P_1$ ($P_2$) 
three-momentum vector, $\vec{k_1}(\vec{q_1})$ in the $V_1 ( V_2)$ rest 
frame and the direction of total $V_1$ ($V_2$) three-momentum vector 
defined in the $B$ rest frame.  $\phi$ is the angle between the normals 
to the planes defined by $P_1 P_1^\prime$ and $P_2 P_2^\prime$, in the 
$B$ rest frame.

The differential decay rate is then given by \cite{Krammer,Kamal},
\begin{eqnarray}
\frac{d\Gamma}{d\cos\theta_1 d\cos\theta_2 d\phi}&=&\displaystyle
N\Bigl(\displaystyle |A^0|^2\cos^2\theta_1 \cos^2\theta_2 
+\frac{|A^\perp|^2}{2}\sin^2\theta_1 
\sin^2\theta_2\sin^2\phi\nonumber  \\ &&+\frac{|A^{||}|^2}{2}\sin^2\theta_1 
\sin^2\theta_2\cos^2\phi +\frac{\Re(A^0 A^{|| *})}{2 \sqrt{2}} \sin 
2\theta_1 \sin 2\theta_2\cos\phi \nonumber \\ &&  -
\frac{\Im(A^\perp A^{0 *})}{2 \sqrt{2}} \sin 2\theta_1 \sin 
2\theta_2\sin\phi-\frac{\Im(A^\perp A^{\| *})}{2 } \sin^2\theta_1 \sin^2 
\theta_2\sin2\phi\Bigr),
\label{dgamma}
\end{eqnarray}
where, $N=\displaystyle\frac{|\vec{k}|}{16 \pi^2 M^2}\frac{9}{4} Br(D^*\to D \pi)$.
The rich kinematics of the vector-vector final state, allows
separation of each of the six combinations of the linear polarization
amplitudes, in the above. Using Fourier transform in $\phi$ and orthonormality of 
Legendre Polynomials in $\cos\theta_1(\cos\theta_2)$, it is possible to 
construct weight functions that project out each of these six 
combinations. An observable ${\cal O}_i$ can then be determined from its 
weight factor ${\cal W}_i$ given in Table~\ref{TI}, using 
\[{\cal O}_i=\int d\cos\theta_1 d\cos\theta_2 d\phi\;\frac{{\cal W}_i}{N} 
\frac{d\Gamma}{d\cos\theta_1 d\cos\theta_2 d\phi}.\] The weight
functions in Table I are not unique and they can be optimized through
numerical simulations.  No additional measurements are required in the
determination of these observables, as the reconstruction of the
vector-vector modes itself generates the angular distributions
required. 

We first focus our attention on the case of a  charged B meson decaying 
to $D^*V$, $V\!\in\!\{K^*,\rho\}$. These final states involve only tree 
level amplitudes and no penguin contributions. The amplitude for the 
$B^+$ decays for a given linear polarization state `$\lambda$' can be written 
as
\begin{equation}
\begin{array}{ll}
\dsp A^\lambda(B^+\to D^{*0} V^+)= V_{ub}^* V_{cq} {\cal A}_u^\lambda e^{i\delta^\lambda_u},
& A^\lambda(B^+\to \overline{D^{*0}} V^+)= V_{cb}^* V_{uq} {\cal A}_c^\lambda 
 e^{i\delta^\lambda_c}   \\[1ex]
\end{array}
\end{equation}
where $q=s$ for $V=K^*$ and $q=d$ for $V=\rho$;
 $\lambda=\{0,\|,\perp\}$. It may be noted that ${\cal A}_u^\lambda$
 and ${\cal A}_c^\lambda$ are real. Since, $D^{*0}$ and
 $\overline{D^{*0}}$ belong to different isodoublets, ${\cal
 A}_u^\lambda$ and ${\cal A}_c^\lambda$ as well as the corresponding
 strong phases $\delta^\lambda_u$ and $\delta^\lambda_c$ are not
 related. {\em No assumption is made regarding the explicit form of the
 amplitudes $A_{c,u}^\lambda$ or the strong phases
 $\delta_{c,u}^\lambda$}. For instance, the amplitudes
 $A_{c,u}^\lambda$ could include contributions from $W$--exchange and
 annihilation diagrams as well, since these involve the same CKM
 phases. Further, our approach does not require the use of
 factorization approximation.  The amplitude for the anti-particle
 decay, $A^\lambda(\overline{B}\to\overline {D^*}\;\overline{V}$) has
 the same strong phases but opposite weak phases to that of
 $A^\lambda(B\to D^* V$). In addition using CPT invariance, for the
 $B^-$ decays we get
\begin{equation}
\begin{array}{ll}
\dsp A^\lambda(B^-\to D^{*0} V^-)= \sigma^\lambda V_{cb} V_{uq}^* {\cal A}_c^\lambda 
e^{i\delta^\lambda_c}, &
 A^\lambda(B^-\to \overline{D^{*0}} V^-)= \sigma^\lambda  V_{ub} V_{cq}^*  {\cal A}_u^\lambda 
 e^{i\delta^\lambda_u} \\[1ex]
\end{array}
\end{equation}
where, $\sigma^\perp=-1, \sigma^{0,\|}=1$.

We consider $D^{*0}/\overline{D^{*0}}$ decaying into $D^0\pi^0 / 
\overline{D^0}\pi^0$, with $D^0/\overline{D^0}$ meson further decaying 
to a final state `$f$' that is common to both $D^0$ and 
$\overline{D^0}$. $f$ is chosen to be a Cabibbo allowed mode of $D^0$ 
(hence, doubly suppressed mode of $\overline{D^0}$). To be specific we may 
take $f=K^- \rho^+$, as this  has the largest branching ratio among 
two--body hadronic decay modes, $Br(D^0\to K^- \rho^+)\approx
10.8\%$\cite{PDG}. 
The accompanying $V$, decays to $K\pi$ for $V=K^*$ and to $\pi\pi$ for 
$V=\rho$. In the $D^{0}-\overline{D^{0}}$ system, CKM predicts 
negligible mixing effects, which we disregard. The amplitudes for the 
decays of $B^+,B^-$ to a final state  involving $f$ and its $CP$ 
conjugate, will be a sum of the contributions from $D^{*0}$ and
$\overline{D^{*0}}$ and can be written as, 
\begin{eqnarray}
A^\lambda_f&=&A^\lambda(\dsp B^+\!\to [\;[f]_{_D}\pi]_{_{D^*}}V^+)=
 \sqrt{B} ( V_{ub}^*V_{cq}  {\cal A}_u^\lambda e^{i\delta^\lambda_u}+
   V_{cb}^* V_{uq}{\cal R} {\cal A}_c^\lambda  e^{i\delta^\lambda_c} e^{i\Delta})\nonumber \\
\bar{A^\lambda_{\bar{f}}}&=&A^\lambda(\dsp B^-\!\to 
[\;[\overline{f}]_{_D}\pi]_{_{D^*}}V^-)=
\sigma^\lambda \sqrt{B} (  V_{ub} V_{cq}^*  {\cal A}_u^\lambda 
  e^{i\delta^\lambda_u}+V_{cb} V_{uq}^* {\cal R} {\cal A}_c^\lambda e^{i\delta^\lambda_c} 
  e^{i\Delta})
\nonumber \\
\bar{A^\lambda_f}&=&A^\lambda(\dsp B^-\!\to [\;[f]_{_D}\pi]_{_{D^*}}V^-)=
\sigma^\lambda \sqrt{B} (V_{ub} V_{cq}^* {\cal R}  {\cal A}_u^\lambda 
 e^{i\delta^\lambda_u} e^{i\Delta} + V_{cb} V_{uq}^* {\cal A}_c^\lambda
e^{i\delta^\lambda_c} )
\nonumber \\ 
A^\lambda_{\bar{f}}&=&A^\lambda(\dsp B^+\!\to [\;[\overline{f}]_{_D}\pi]_{_{D^*}}V^+)=
\sqrt{B} ( V_{ub}^*
V_{cq}  {\cal R} {\cal A}_u^\lambda e^{i\delta^\lambda_u} e^{i\Delta}+
 V_{cb}^* V_{uq} {\cal A}_c^\lambda 
 e^{i\delta^\lambda_c}),
\label{amps}
\end{eqnarray}
where, $[X]_{_M}$ indicates that the state $X$ is reconstructed to
have the invariant mass of $M$; $B=Br(D^0\to f), {\cal
R}^2=Br(\overline {D^{0}}\to f)/Br(D^{0}\to f)$ and $\Delta$ is the
strong phase difference between $D^0\to f$ and $D^0\to {\overline f}$
(or that between $D^0\to f$ and ${\overline D^0\to f}$, since $D^0\to
{\overline f}$ and ${\overline D^0 \to f}$ have the same
strong phase).

A measurement of the angular distribution given in eqn.(\ref{dgamma}),
for each of the four modes noted above in (\ref{amps}), yield a total
of twentyfour observables, six for each mode.  These can be extracted
experimentally using Table~\ref{TI}. This is much larger than the
sixteen unkowns: $R,\Delta,\gamma,|V_{ub}|$ and three variables for
each of, $A_u^\lambda,A_c^\lambda,\delta^\lambda_u$, and
$\delta^\lambda_c$. Thus, $\gamma$ would be over-determined and sign
ambiguities possibly resolved. Since, $|V_{ub} V_{cq}^*|{\cal R} {\cal
A}_u^\lambda \ll |V_{cb}^* V_{uq}|{\cal A}_c^\lambda$, the last two
equations in eqn.(\ref{amps}), may not be distinguishable, i.e.,
$|\bar{A^\lambda_f}| \approx |A^\lambda_{\bar{f}}|$ . This reduces the
number of independent equations to eighteen, but still allows $\gamma$
to be determined.  The conditions, ${\cal R},\frac{{\cal A}_u^\lambda}
{{\cal A}_c^\lambda}\ll 1$, can also help reduce the sign ambiguities.

It is well known that a study of the angular correlations can be used
to extract $CP$ violating asymmetries\cite{DQSTL}.
In addition to the usual signature of $CP$ violation,
\begin{equation} 
| A_f^\lambda|^2-|\bar{A^\lambda}_{\bar{f}}|^2=4 |V_{ub}^* V_{cq} V_{cb} V_{uq}^*|{\cal R} B 
{\cal A}_u^\lambda{\cal A}_c^\lambda  \sin(\delta^\lambda_c-\delta^\lambda_u+\Delta) \sin\gamma,
\label{usualass}
\end{equation}
the complete study of the angular distribution of vector-vector final
states, provides the following alternative signatures for $CP$ violation,
\begin{eqnarray}
\Im\{(A^\lambda A^{\rho*})_f &&+(\bar{A^\lambda}\bar{A^{\rho*}})_{\bar f} \nonumber \\
&&=2 |V_{ub}^* V_{cq} V_{cb} V_{uq}^*| {\cal R} 
B \sin\gamma
\left( {\cal A}_u^\lambda {\cal A}_c^\rho \cos(\delta_u^\lambda-\delta_c^\rho-\Delta)- 
{\cal A}_c^\lambda {\cal A}_u^\rho
\cos(\delta_c^\lambda-\delta_u^\rho+\Delta )\right)
\label{ass1}
 \\
\Im\{(\bar{A^\lambda}\bar{A^{\rho*}})_f &&+(A^\lambda A^{\rho*})_{\bar f}\} \nonumber \\
&&=2 |V_{ub}^* V_{cq} V_{cb} V_{uq}^*| {\cal R} 
B \sin\gamma
\left( {\cal A}_u^\lambda {\cal A}_c^\rho \cos(\delta_u^\lambda-\delta_c^\rho+\Delta)- 
{\cal A}_c^\lambda {\cal A}_u^\rho
\cos(\delta_c^\lambda-\delta_u^\rho-\Delta )\right)\label{ass2} \\
\Im\{(A^\lambda A^{\rho*})_f &&+(\bar{A^\lambda}\bar{A^{\rho*}})_{\bar f}+
(\bar{A^\lambda}\bar{A^{\rho*}})_f+(A^\lambda A^{\rho*})_{\bar f}\} \nonumber \\
&&=4 |V_{ub}^* V_{cq} V_{cb} V_{uq}^*| {\cal R} 
B \sin\gamma\cos\Delta
\left( {\cal A}_u^\lambda {\cal A}_c^\rho \cos(\delta_u^\lambda-\delta_c^\rho)- {\cal A}_c^\lambda {\cal A}_u^\rho
\cos(\delta_c^\lambda-\delta_u^\rho )\right)
\label{ass3}
\end{eqnarray}
where $\lambda=\perp$, $\rho=\|$ or $0$. The signals in
eqns.(\ref{ass1})-(\ref{ass3}) are coefficients of $\sin \phi$ and
$\sin 2\phi$ in the angular distribution in eqn.(\ref{dgamma}).  The
advantage here is that these signals of $CP$ violation are not diluted
by sine of strong phase as was the case in eqn(\ref{usualass}) and
also that, they are obtained by adding $B$ and $\overline{B}$ events.
We wish to emphasize that, the angle $\phi$ between the planes of the
decay products of $D^*$ and V, plays a crucial role. If one measures
only
$|A^\lambda_f|^2,|\bar{A^\lambda}_{\bar{f}}|^2,|\bar{A^\lambda}_f|^2$
and $|A^\lambda_{\bar{f}}|^2$, and were to overlook the interference
terms of the helicity amplitudes that appear in the complete angular
distribution, one would have twelve observables by considering all
three polarizations with a total of thirteen unknowns. Unless, one of
the variables is assumed to be measured elsewhere, $\gamma$ cannot be
extracted. The situation would be worse if $|\bar{A^\lambda_f}|
\approx |A^\lambda_{\bar{f}}|$, as there would be even fewer
observables than unknowns, rendering $\gamma$ truly unmeasurable if
the $A^\lambda A^{\rho*} (\lambda\neq\rho)$ terms are ignored.

Note that since all the amplitudes and strong phases involved in the  
right-hand-side of eqn.(\ref{amps}) are solved for, using the
observables constructed from these amplitudes, we need not disentangle 
the strong phases associated with each isospin state of the various 
partial waves. 

In the case of neutral B mesons the $D^*K^*$ decay mode is self
tagging\cite{Dunietz} if $K^{*0}/\overline{K^{*0}}$ is seen in the
$K^+\pi^-/K^-\pi^+$ mode. Hence, no time dependent measurements are
required and the observables for the decays of $B^0$ and
$\overline{B^0}$ to any final state `$f$' and its $CP$ conjugate, may
be obtained by the replacement of the charged B decay amplitudes
$A^\lambda_{u,c}$ by the corresponding neutral B amplitudes
$a^\lambda_{u,c}$ in eqn.(\ref{amps}). Within factorization
approximation, $a_c$ differs from $A_c$, due to the fact that the
charged $B$ decay amplitudes include contributions from both color
allowed as well as color suppressed diagrams, whereas neutral $B$
decay amplitudes come only from the color suppressed diagrams; $a_u$
and $A_u$, however, are identical. The signatures of $CP$ violation
are similar to eqns.(\ref{ass1})-(\ref{ass3}), with $A_{c,u}$ replaced
by $a_{c,u}$. Even in the case, where tagging is not possible, $B^0$
and ${\overline B^0}$ observables can be added resulting in an
asymmetry independent of the mixing parameters, $\Delta m/\Gamma$ and
$\beta$, and again of the same form as in
eqns.(\ref{ass1})-(\ref{ass3}).  Addition of $B^0$ and ${\overline
B^0}$ observables reduces the number of available equations and hence,
we need to consider $D^0/ {\overline D^0}$ decaying not only the final
state `$f$' but also an additional $CP$--eigenstate. Further, all
three linear polarization states will have to be analyzed. {\em This
makes it possible to extract $\gamma$ without any need for time or
flavor tagging}.

Next, we construct $CP$ violating asymmetries corresponding to the
signals suggested in eqn.(\ref{ass1}). As pointed out earlier, the
coefficients of the $\sin \phi$ and $\sin 2 \phi$ terms need to be
isolated, in order to obtain $\Im(A^\perp A^{0*})$ and $\Im(A^\perp
A^{\|*})$ terms, respectively.  The coefficient of the $\sin \phi$
term in eqn.(\ref{dgamma}), can be determined by defining the
following asymmetry,
\[
\displaystyle A_1= \frac{\displaystyle (\displaystyle
  \int_0^\pi-\int_\pi^{2\pi})
 d\phi\;\displaystyle\int_D
 d\cos\theta_1\; \int_Dd\cos\theta_2 \;\frac{d\Gamma_{sum}}{d\cos\theta_1
d\cos\theta_2 d\phi}} {
 \displaystyle
\int_0^{2\pi}  
d\phi\;\displaystyle \int_S
 d\cos\theta_1\; \int_S d\cos\theta_2 \;\frac{d\Gamma_{sum}}{d\cos\theta_1
d\cos\theta_2 d\phi}}\;,
\]
where
$\displaystyle\int_{D(S)}\equiv\displaystyle\int_{-1}^0\mp\int_0^{1}$ and
$\Gamma_{sum}=\displaystyle\Bigl(\Gamma(B\!\to 
[\;[f]_{_D}\pi]_{_{D^*}}V)+\Gamma(\overline{B}\!\to 
[\;[\bar{f}]_{_D}\pi]_{_{D^*}}\overline{V})\Bigr)$. On performing the angular integrals this 
asymmetry is equivalent to,
\begin{equation}  
\displaystyle A_1=\:\frac{-2 \sqrt{2}}{\pi} \frac{\displaystyle 
\Im\{(A^\perp A^{0*})_f +(\bar{A^\perp}\bar{A^{0*}})_{\bar f}\}}
{ \displaystyle \sum_{\lambda=\perp,\|,0}(| A_f^\lambda|^2+|\bar{A^\lambda}_{\bar{f}}|^2)}.
\nonumber\end{equation}
Yet another symmetry comes from the coefficient of the $\sin 2\phi$ term in
eqn.(\ref{dgamma})  and is defined as, 
\begin{eqnarray}  
\displaystyle A_2 &=& \frac{\displaystyle 
 (\displaystyle\int_0^{\pi/2}-\int_{\pi/2}^{\pi}+\int_{\pi}^{3\pi/2}
-\int_{3\pi/2}^{2 \pi}) d\phi\;\displaystyle\int_S d\cos\theta_1\;
\int_S d\cos\theta_2 
 \;\frac{d\Gamma_{sum}}{d\cos\theta_1 d\cos\theta_2 d\phi}}
{ \displaystyle
\int_0^{2\pi} d\phi\;\displaystyle
\int_S d\cos\theta_1\;
\int_S d\cos\theta_2 \;\frac{d\Gamma_{sum}}{d\cos\theta_1 d\cos\theta_2 
d\phi}}\;,\nonumber\\
\displaystyle  &=&\:\frac{-4}{~\pi} \frac{\displaystyle 
\Im\{(A^\perp A^{\|*})_f +(\bar{A^\perp}\bar{A^{\|*}})_{\bar f}\}}
{ \displaystyle \sum_{\lambda=\perp,\|,0}(|
 A_f^\lambda|^2+|\bar{A^\lambda}_{\bar{f}}|^2)}.
\end{eqnarray}
The asymmetries $A_1$ and $A_2$ can be  similarly constructed for the
signals in eqns.(\ref{ass2})-(\ref{ass3}). However, these will be 
much smaller as they involve interference of amplitudes that are not 
comparable. 

We now compute a rough estimate of the number of $B$'s required to
observe the $CP$ violating signal in our method.  Exact numbers can of
course only be obtained, once the strong phases $\delta^\lambda_{u,c}$
and $\Delta$, as well as the amplitudes $A^\lambda_{u,c}$, are
determined from the observables measured experimentally.  For our
estimates, we set $\delta^\lambda_{u,c}=\Delta=0$ and $|V_{ub}
V_{cs}/({V_{cb} V_{us}})|=0.38$. The form-factors and decay constants
are chosen from Ref.~\cite{BSW} (which uses the factorization
approximation) and the ratio of the coefficients of 
color supressed ($\sim a_2$ ) to color allowed ($\sim a_1$) amplitudes
(as defined in Ref.~\cite{BSW}) is taken to be $|a_2/a_1|\approx
0.26$\cite{Browder}.  $\cal R$ is esimated as \cite{ADS,CLEO},
\[ {\cal R}^2=\frac{Br(\overline{D^0}\to K^-\pi^+)}{Br(D^0\to K^-\pi^+)}=
\frac{Br(\overline{D^0}\to K^-\rho^+)}{Br(D^0\to K^-\rho^+)}=0.0077 .
\]
The resulting asymmetries for $B^+\to D^* K^{*+}$($B^+\to D^* \rho^+$)
at $\gamma=\displaystyle\frac{\pi}{6}$ are found to be
$A_1=-28\%(0.5\%),\:A_2=9.1\%(-0.16\%)$. The total number of charged
$B$'s required to observe these asymmetries at $3\sigma$ significance
are $N_1^{3\sigma}=7.6\times 10^8(4.2\times 10^{10}),\:
N_2^{3\sigma}=7.3\times 10^9(4.0\times 10^{11})$. The number of $B$'s
required can easily be reduced by a factor of $\sim 3 - 4$, if one
sums over all the doubly Cabbibo suppressed modes of $\overline{D^0}$.
The corresponding asymmetries for the neutral $B$'s vanish
identically, under the factorization approximation, in the absence of
strong phases. This is due to the fact that the factorization
approximation implies $a^\lambda_u=a^\lambda_c$. A test of this
relation would provide a unique model independent test of the
factorization approximation.

To conclude, we have extended the GLW and ADS proposals
to measure $\gamma$ using vector-vector final states. The rich
kinematics of these modes provide a large number of observables that
can be obtained using appropriate weight functions, if the angular
distributions are available. The reconstruction of these modes, itself
generates the angular distributions required. One particular final
state is enough to extract $\gamma$ as well as all the hadronic
amplitudes and strong phases involved.

We are extremely grateful to Prof. L. M. Sehgal for detailed
discussions and valuable suggestions. We also thank
Prof. G. Rajasekaran for discussions and encouragement.

\begin{table}
\caption{The weight factors corresponding to the observables in the
angular distribution (eqn.(\protect\ref{dgamma})) for $B\to VV$ decays. Note
that the weight factors would give identical results under $\theta_1\leftrightarrow\theta_2$.}
\label{TI}
\begin{center}
\begin{tabular}{ccc}
~~~~~~~~~~~~~~~~~~~~Observable ${\cal O}_i$~~~~~~~~~~~~~~~~~~~~& weight 
${\cal W}_i$\\
\hline
$|A^0|^2 $         & $\displaystyle\frac{3}{16 \pi}(15 \cos^2\theta_1-3)$  \\[1ex]
$|A^{||}|^2$       & $\displaystyle\frac{3}{16 \pi}(-6+12 \cos^2\phi
                       +\frac{9-15 \cos^2\theta_1}{2})$    \\[2ex] 
$|A^\perp|^2$      & $\displaystyle\frac{3}{16 \pi}(6-12 \cos^2\phi
                       +\frac{9-15 \cos^2\theta_1}{2})$    \\[2ex] 
$\Re(A^0 A^{|| *})$ &$\displaystyle\frac{2^5\sqrt{2}}{\pi^3}\cos\phi
                       \cos\theta_1\cos\theta_2$  \\[2ex]
$\Im(A^\perp A^{0 *})$ &$\displaystyle-\frac{2^5\sqrt{2}}{\pi^3}\sin\phi
                       \cos\theta_1\cos\theta_2$      \\[2ex]
$\Im(A^\perp A^{\| *})$ &$\displaystyle-\frac{9}{8\pi}\sin2\phi$ 
\end{tabular}      
\end{center}
\end{table}

\end{document}